\documentclass[prl,superscriptaddress,onecolumn]{revtex4}

\pdfoutput=1

\usepackage{graphicx}
\usepackage{amssymb,amsmath}
\usepackage{subfigure}
\newcommand{\dd}{\mathrm{d}}

\begin{document}

\author{Philipp~Werner*}
\affiliation{Theoretische Physik, ETH Zurich, 8093 Z{\"u}rich, Switzerland}
\author{Michele~Casula}
\affiliation{CNRS and Institut de Min\'eralogie et 
de Physique des Milieux condens\'es,
case 115, 4 place Jussieu, 75252, Paris cedex 05, France}
\author{Takashi~Miyake}
\affiliation{Nanosystem Research Institute (NRI) 
RICS, 
AIST, Tsukuba, Ibaraki 305-8568, Japan }
\affiliation{Japan Science and Technology Agency, CREST}
\affiliation{Japan Science and Technology Agency, TRIP}
\author{Ferdi~Aryasetiawan}
\affiliation{Graduate School of Advanced Integration Science, Chiba University, 
Chiba 263-8522, Japan}
\author{Andrew~J.~Millis}
\affiliation{Department of Physics, Columbia University, 538 West, 120th Street, New York, NY 10027, USA}
\author{Silke~Biermann}
\affiliation{Centre de Physique Th{\'e}orique,
Ecole Polytechnique, CNRS-UMR7644, 91128 Palaiseau, France.}
\affiliation{Japan Science and Technology Agency, CREST}

\title{
Satellites and large doping- and temperature-dependence of electronic properties in hole-doped BaFe$_2$As$_2$
}

\maketitle

{\bf
Over the last years, superconductivity has been
discovered in several families of iron-based compounds.
Despite intense research, even basic electronic
properties of these materials, such as Fermi surfaces, 
effective electron masses, or orbital characters
are still subject to debate.
Here, we address an issue that has not been
considered before, namely the consequences of dynamical
screening of the Coulomb interactions among Fe-$d$ electrons.
We demonstrate its importance not only for correlation 
satellites seen in photoemission spectroscopy, 
but also for the low-energy electronic structure.
From our analysis of the normal phase of BaFe$_2$As$_2$ 
emerges the picture of a strongly correlated compound
with strongly doping- and temperature-dependent properties.
In the hole overdoped regime, an incoherent metal is found, while 
Fermi-liquid behavior is recovered in the undoped compound. 
At optimal doping, the self-energy exhibits an unusual square-root 
energy dependence which leads to strong band renormalizations 
near the Fermi level. 
}

Most known superconductors can be attributed to one of two classes:
The first class comprises materials such as MgB$_2$, classified
as ``weakly correlated'' in the sense that one-electron 
theories work well for the description of the
basic electronic properties, and the superconducting pairing
mechanism has eventually been understood in this framework.
Materials of the second class exhibit sometimes spectacular 
failures of the one-electron picture, leaving unclear even
the theoretical language in which a theory of the pairing
mechanism should be formulated.
High-T$_c$ cuprate superconductors fall in the latter category.
The parent compounds of these materials are Mott insulators due to
strong Coulomb interactions that localize the electrons on the Cu sites.
The role of electronic correlations in the recently 
discovered iron-based high-T$_c$ 
superconductors \cite{Kamihara08}
is less clear, and apparently dependent on the specific family, 
as well as on doping, substitutions or pressure.
In this work, we address the prototypical compound
of the so-called ``122-family'',
BaFe$_2$As$_2$, which exhibits superconductivity under pressure
\cite{alireza,kimber} or hole- as well as electron-doping \cite{rotter,sefat}.
Various experimental probes -- angle-resolved and angle-integrated
photoemission spectroscopy 
\cite{liu,brouet,Ding08,Borisenko09,Fink10, Zhang11, Mansart11}, 
optics and transport, 
Raman spectroscopy,
neutron experiments, NMR --
have been employed to characterise the electronic
properties \cite{Wen11}. The Fermi surface consists of two concentric
hole pockets
around the $\Gamma$ point, and elliptic 
electron pockets around M \cite{singh}.
Experimental estimates of the (doping-dependent) mass enhancements vary
from about $1.4$ \cite{Yi09} to $5$, at least for the 
orbital pointing towards the As-sites \cite{brouet-arxiv}.  
The orbital character of these pockets are still subject to
debate, but there seems to emerge a consensus
about stronger correlation effects for 
holes than electrons.

Understanding the low energy electronic
structure, in particular Fermi surface nesting, orbital character
and mass enhancements, is a prerequisite for assessing possible
pairing mechanisms \cite{ding,kuroki08,kontani,Wen11}.
While the field of electronic structure calculations
for correlated materials has made tremendous progress
in recent years, most notably thanks to the combination of 
electronic structure and many-body techniques, the
discovery of iron pnictides provides a serious
challenge.
The combined ``LDA+DMFT'' scheme builds
on density functional theory within
the local density approximation (LDA) to construct
realistic many-body Hamiltonians, which are solved using 
dynamical mean field theory (DMFT) \cite{Georges96,Anisimov97,Lichtenstein98}.
While a number of interesting applications of this scheme
to iron pnictides  
(see e.g. \cite{shim,anisimov,aichhorn09})
have emerged,
it has become clear that one of the bottlenecks is the determination of 
the Coulomb matrix elements (``Hubbard parameters'' $U$ and $J$),
that parametrize the energetic cost associated with the distribution of electrons
among the localized Fe-$d$ orbitals.
Recent works try to extract these values from 
constrained LDA calculations \cite{anisimov}, 
GW-inspired methods \cite{kupetov},
or from random-phase
approximation (RPA) based schemes.
The constrained RPA approach \cite{aryasetiawan04} 
takes into account the dynamical nature of screening, 
so that the ``Hubbard $U$" for the Fe-$d$ states 
becomes a frequency-dependent object. 
None of the theoretical studies so far
have treated dynamical Coulomb interactions 
at the level of the actual many-body calculation.
Here, we demonstrate the importance of
the frequency-dependence for the
low-energy electronic structure, 
mass enhancements and lifetimes, as well as for
a description of correlation satellites
seen in photoemission spectroscopy.

\begin{figure}[h]
\centering
\includegraphics[angle=0, width=\linewidth]{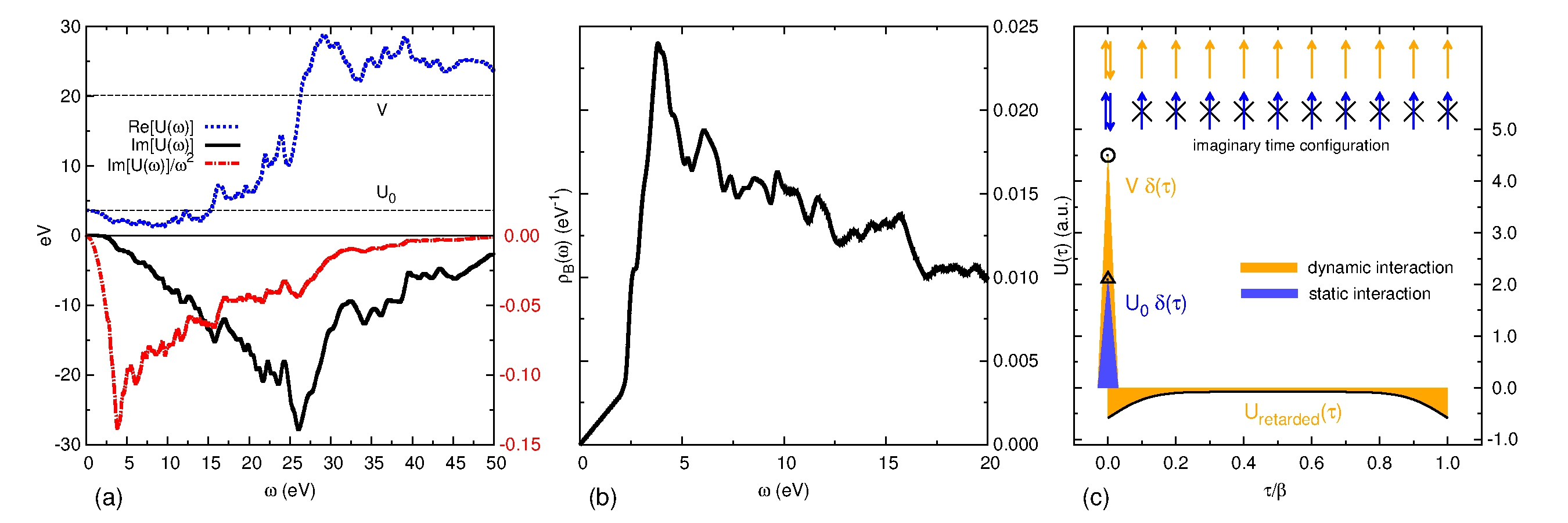}
\caption{Frequency dependent interaction for BaFe$_2$As$_2$ from constrained RPA. Panel (a) shows the real and imaginary parts of the average intra-orbital Coulomb interaction as a function of frequency. $V$ is the unscreened (bare) Coulomb interaction and $U_0$ the static (partially screened) interaction. $\text{Im}U(\omega)/\omega^2$, plotted as a red line, features prominent peaks at 3.8 eV, 6.1 eV, 16 eV, and 26 eV, and smaller humps at 10 eV and 12 eV. Panel (b) shows the spectral function of the bosonic propagator $B$. The peaks of this function, inherited from the structures in $\text{Im}U(\omega)/\omega^2$, determine the energies of satellite structures in the $d$-electron density of states. A schematic plot illustrating the difference between a static-$U$ (yellow) and a dynamic-$U$ (blue) calculation is shown in panel (c). In a static-$U$ calculation, up-spin and down-spin electrons interact locally in time with the partially screened Coulomb interaction $U_0$. In the dynamic-$U$ calculation, the instantaneous interaction is given by the bare $V$, while screening effects lead to an attractive retarded interaction $U_\text{retarded}(\tau)$.}
\label{ufreq}
\end{figure}

\noindent \textbf{Partially screened Coulomb interaction}

The constrained RPA result for the average intra-orbital Coulomb repulsion $U$ 
of BaFe$_2$As$_2$ is shown in Fig.~\ref{ufreq}a. 
The real part ranges from the static value $U_0\equiv \text{Re}U(\omega=0)=3.6$~eV 
to the bare interaction $V$ of about $20$~eV at large $\omega$, and the frequency dependence
resembles typical $U(\omega)$ in transition metals \cite{aryasetiawan04, miyake08}, 
where there is no well-defined plasmon excitation.
Instead, the main high-energy excitation with a peak in $\text{Im}U(\omega)$ at $\sim 26$~eV forms a
broad structure extending down to a few eV, implying that the plasmon
excitation overlaps strongly with the one-particle excitations, a reflection
of the semi-itinerant nature of the Fe-3$d$ electrons.
$U(\omega)$ represents a {\it partially screened} Coulomb interaction for the Fe-$d$ states, which accounts for screening
by all degrees of freedom except the Fe-$d$ states themselves. Fourier transformation yields
a Hubbard interaction term 
$\int\int d\tau d\tau^{\prime} N(\tau)U(\tau - \tau^{\prime})N(\tau^{\prime})$, 
with $U(\tau)=V\delta(\tau)+U_\text{retarded}(\tau)$, 
which describes the local Coulomb
interaction including retardation effects due to quantum fluctuations that screen the bare $V$
(see Fig.~\ref{ufreq}c and Methods Section).

\begin{figure}[h]
\centering
\includegraphics[width=\linewidth]{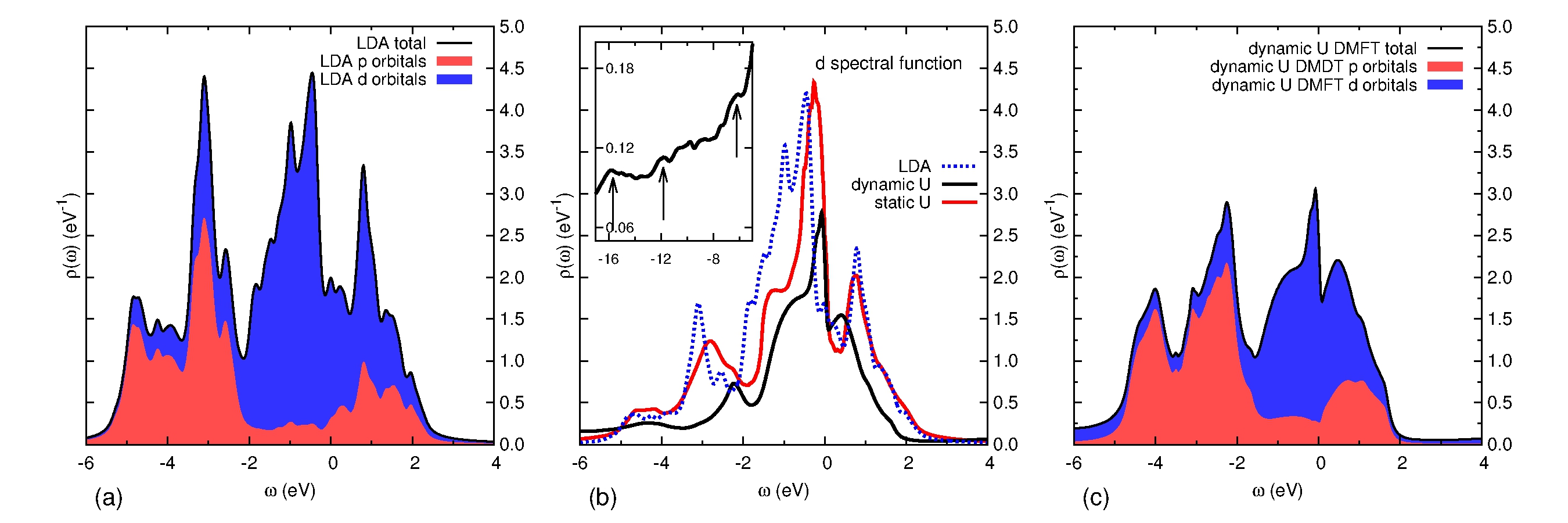}
\caption{Effect of $U(\omega)$ on the $d$-electron spectral function. Panel (a) shows the total LDA density of states for BaFe$_2$As$_2$ with the $p$- and $d$-electron contribution shaded in red and blue, respectively. Panel (b) shows the $d$-electron spectral function obtained from the DMFT calculation at $T=290$ K and optimal hole doping ($x=0.2$ per Fe), using the static value $U_0$ (red curve) and the full frequency dependent $U(\omega)$ (black curve). The inset shows the high-energy tail of the $d$-electron spectral function. Arrows indicate the position of satellites predicted by the dynamic-$U$ calculation. In panel (c), we plot the total spectral function ($p$- and $d$-contribution) of optimally doped BaFe$_2$As$_2$ obtained from DMFT.}
\label{spectra}
\end{figure}

In a standard LDA+DMFT calculation without dynamical
effects, the interaction $U_0$ would result in a rather weakly correlated picture.
This is demonstrated in Fig.~\ref{spectra}b, where the Fe-$d$ spectral 
function from such a static-$U$ LDA+DMFT calculation is shown. Interaction effects 
lead to a moderate renormalization of the Fe-$d$ states, with
a mass enhancement of $1.6$. 
A comparison to the LDA density of states in panel (a) shows that the peaks at 
$-3$~eV and $1$~eV are weakly renormalized band states. 
No Hubbard satellites or other correlation features appear, in agreement with previous
studies \cite{anisimov, yin2011}.

The new aspect of our work, compared to previous simulations, 
is the treatment
of the full frequency dependence of the interaction. In Refs.~\onlinecite{Werner07, Werner10} 
it was shown in the context of simple model
calculations, how a frequency dependent interaction can be incorporated
efficiently into Quantum Monte Carlo simulations within
the hybridization expansion impurity solver scheme \cite{Werner06}.
Here, we generalize this technique to multi-orbital systems and 
realistic materials. 
The effect of $U(\omega)$ is to dress the Fermionic
propagators with a bosonic propagator
\begin{equation}
B(\tau)=\exp\left[-K(\tau)\right],\label{bose_definition}
\end{equation}
where $K(\tau)$ is the twice integrated retarded interaction.
In terms of $\text{Im}U(\omega)$ 
and a factor $b(\tau,\omega)=\cosh \big[\big(\tau -\frac{\beta }{2}\big)\omega\big]/\sinh \big[\frac{\beta \omega}{2}\big]$ with bosonic symmetry,
we can write \cite{Werner10}
\begin{equation}
K(\tau )=\int_{0}^{\infty }\frac{d\omega}{\pi }\frac{\text{Im}U(\omega)}{\omega^2}[b(\tau,\omega)-b(0,\omega)].
\label{k_definition}
\end{equation}
It is evident from Eq.~(\ref{k_definition})
that, as far as structures in
the frequency dependent interaction are concerned, 
the relevant function to analyze
is $\frac{\text{Im}U(\omega )}{\omega ^{2}}$. 
This function is 
plotted as red dash-dotted line in Fig.~\ref{ufreq}a. Besides a first peak at 
$3.8$~eV, which comes from the rapid decay of $\text{Im}U(\omega )$ at
small frequencies, there are prominent peaks at $6.1$~eV, $16$~eV and $26$~eV, as well as
smaller features at $10$~eV and $12$~eV. 
We have traced the
origin of the $6.1$~eV feature to transitions from occupied $d$-states 
to states in the energy window [$6$~eV:$7$~eV], which have predominantly Ba 
character.
The $12$ eV feature is due to transitions from As-$s$ semicore states
to empty Fe-$d$ states close to the Fermi level.

\noindent \textbf{Density of states and high-energy satellites}

An interesting question is how the structures in $U(\omega)$ 
affect the properties of the material.
As discussed in Ref.~\onlinecite{dyn_anal_cont} and in the Supplementary Material, 
the Green function for a system with frequency dependent interactions $U(\omega)$ 
factorizes {\it in the atomic limit} into the atomic Green function for the system with static
interactions $U_0$ and a bosonic factor, 
which is precisely the propagator $B$ defined in Eqs.~(\ref{bose_definition}) and (\ref{k_definition}).
Inspired by this atomic limit property, we introduce the auxiliary Green 
function $G_\text{auxiliary}$ by 
$G(\tau) = G_\text{auxiliary}(\tau) B(\tau)$
and compute the spectral function of the system as 
the {\it convolution} of the spectral functions of the
auxiliary propagator, $\rho_\text{aux}(\omega)$, 
and of the
spectral function of the bosonic propagator, $\rho_B(\omega)$  
(see Eq.~(9) in the Supplementary Material).

For systems with a single well-defined plasma frequency
the effect of this convolution would be to replicate the
low-energy spectral features of $\rho_\text{aux}(\omega)$ (with exponentially decreasing
weights), displaced by multiples of the plasma frequency.
These satellites correspond to processes where in
addition to the one-electron addition or removal process
a certain number of plasmons are emitted or absorbed.
In the present case, the bosonic spectrum, plotted in Fig.~\ref{ufreq}b, is more
complex than just a single plasmon delta-function, since 
it inherits the structures of $\text{Im}U(\omega)/\omega^2$.
Nevertheless, sharp features present in $\rho_B(\omega)$ lead to replications of the
structure of the low-energy spectral density.

The low-energy part of the $d$-electron spectral function 
from the dynamic-$U$ calculation is shown in Fig.~\ref{spectra}b, 
and the total spectral function ($p$- and $d$-contribution) in Fig.~\ref{spectra}c. 
In the dynamic-$U$ calculation, spectral weight is
shifted to high energies, 
which leads to a reduction in
weight at low energies, compared to the static-$U$ result. More importantly,
we see that the structures arising from $p$-$d$ hybridization are shifted closer to the Fermi energy, 
and the peak near $\omega=0$ is strongly renormalized. Thus, the explicit treatment of the
strong Coulomb repulsion at large frequencies has a substantial effect, even 
on the {\it low-energy} properties of the system. 
The large quantitative effect of $U(\omega)$ is
remarkable given the fact that the real part of the frequency-dependent
interaction (Fig.~\ref{ufreq}a) remains of the order of $U_0$ or
lower up to $\omega\sim 15$~eV. 

The sharp low-energy peak in the $d$-electron spectral function results
in weak, but well-defined satellites, as discussed above. The inset of Fig.~\ref{spectra}b
shows the high energy tail of the occupied part of the spectrum, with arrows marking
the most prominent satellites at $-6.1$~eV, $-12$~eV and $-16$~eV. This physics is of course
{\it not} contained in a static-$U$ calculation or any other previous theoretical work on pnictides.
The observation of satellites at $-6.5$~eV and $-12$~eV was emphasized in the
photoemission study of Ding and collaborators \cite{Ding08}. While the
feature at $-12$~eV was argued in Ref.~\onlinecite{Jong09} to be an As-4$s$
line, the latter work confirms a hump in the $d$-electron spectral function around $-6.5$~eV. 
Our calculation suggests that a $d$-feature, originating from the structure in the frequency
dependent interaction, is superimposed to the As-4$s$ spectral contribution. 
The $-16$~eV feature is probably not visible in experiments, because it overlaps with Ba-5p states, 
while a satellite which we predict at $-3.8$~eV is masked by structures arising from $p$-$d$ hybridization.

\begin{figure}[h]
\centering
\includegraphics[angle=0, width=\linewidth]{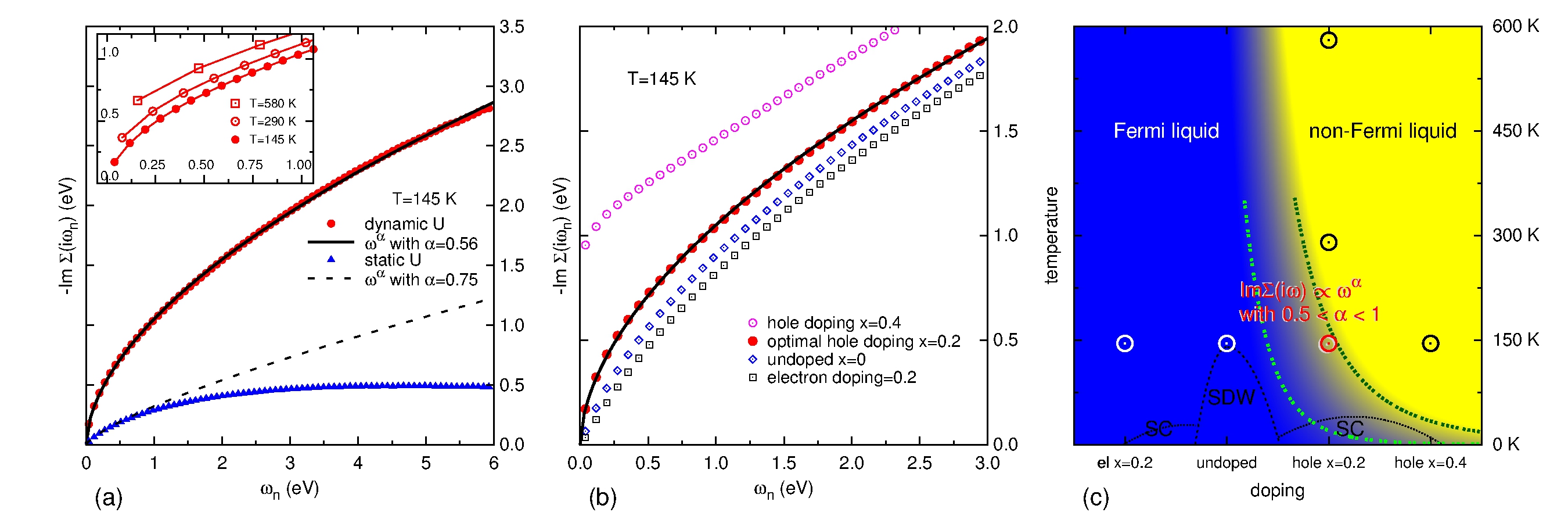}
\caption{Non-Fermi liquid behavior of the self-energy. Panel (a) shows the imaginary part of the self-energy on the Matsubara axis for optimal hole doping ($x=0.2$ per Fe), both for the dynamic-$U$ calculation (red circles) and the static-$U$ calculation (blue diamonds). Solid and dashed lines are fits of the low-frequency behavior to the function $-\text{Im}\Sigma(i\omega_n)=A(\omega_n)^\alpha$. The inset shows the low-frequency behavior of the dynamic-$U$ result for different temperatures. As the temperature is raised, the extrapolation to $\omega_n\rightarrow 0$ yields a non-zero intercept, which indicates  
that even excitations at the Fermi level exhibit a finite lifetime.
Panel (b) shows the low energy behavior of the self-energy as a function of doping. Fermi-liquid behavior is found in the undoped and electron-doped compounds, while a non-zero intercept appears in the overdoped case. 
Panel (c) shows a sketch of the phase diagram in the space of temperature and doping. The blue region indicates Fermi-liquid behavior, while yellow indicates a frequency-dependence of the self-energy which is not compatible with Fermi-liquid theory. The left dashed line marks the boundary of the crossover region, where the exponent $\alpha$ starts to deviate from $1$. The right dotted line corresponds to $\alpha=0.5$, which marks the ``spin freezing" transition. To the right of this line, an incoherent metal phase with non-zero intercept of $\text{Im}\Sigma$ is found. The experimentally measured phase-diagram with superconducting (SC) and spin-density-wave (SDW) ordered phases is indicated by black dotted lines. Full substitution (KFe$_2$As$_2$) corresponds to $x=0.5$.
}
\label{sigma}
\end{figure}

\noindent \textbf{Non-Fermi liquid properties of the metallic phase}

In a Fermi liquid, the imaginary part of the Matsubara axis self-energy
exhibits a linear regime at low energy, whose slope is
directly related to the quasi-particle mass enhancement.
However, as shown in Fig.~\ref{sigma}a, for optimally doped BaFe$_2$As$_2$ ($x=0.2$ hole doping per Fe), 
and at the temperatures of our 
simulations, we do not observe this behavior: the self-energy behaves as
$-\text{Im}\Sigma(i\omega_n)=A(i\omega_n)^\alpha$ with 
$\alpha\approx 0.56$ at $T=145$~K, while a frequency dependence of the form 
$-\text{Im}\Sigma(i\omega_n)=B+A(i\omega_n)^\alpha$, with non-zero intercept 
(scattering rate) $B$ is found at $T=290$~K and $T=580$~K. 

A similar phenomenon has recently been observed in the metallic phase of a 3-orbital model \cite{Werner08}, and has been dubbed 
``spin-freezing" transition, because the intercept $B$ is due to scattering off static (but disordered) local moments. 
The exponent value $\alpha=0.5$ (and simultaneous onset of an intercept $B>0$) was found to mark the transition 
into the spin frozen, or incoherent metal regime. Large deviations from Fermi-liquid behavior thus appear in the vicinity of this transition. 
The phenomena described in Ref.~\onlinecite{Werner08} seem to be a generic
property of the metallic phase in multi-orbital systems with large Hund's coupling.
Our results in Fig.~\ref{sigma}a indicate that optimally doped BaFe$_2$As$_2$, at temperatures somewhat above $T_c$, is close 
to the spin-freezing transition, and that increasing the temperature shifts the material across the spin-freezing boundary into the incoherent 
metal regime. 

A square-root like rather than linear low-frequency behavior of $\text{Im}\Sigma$
means that Landau quasi-particles and effective masses cannot be properly defined. 
The square root implies that the bands very close the the Fermi level are much more strongly renormalized
than those further away, consistent with the behavior seen in Fig.~\ref{spectra}.
This non-Fermi liquid (NFL) property may explain why different photoemission experiments arrive at 
considerably different estimates for band renormalizations,
and why it has been so difficult to reach a consensus on the importance of correlations
in BaFe$_2$As$_2$ (and, possibly, other pnictides).  

While Fermi-liquid properties may eventually be recovered in simulations at low enough temperature (see Fig.~\ref{sigma}c), such behavior is cut off in real materials by the onset of the spin-density-wave or superconducting phase. Thus the NFL and incoherent metal regime dominates the physics in the whole temperature range of relevance to this study. We also note that NFL behavior has been proposed for LaFeAsO and FeSe, based on static-$U$ calculations \cite{Haule_njp,aichhorn,Ishida10}. 

Our result from the static-$U$ calculation is plotted in Fig.~\ref{sigma}a. 
We see that taking into account the frequency dependence leads to a substantially 
enhanced self-energy, and a much larger frequency range over which the NFL exponent 
is valid. Fitting the self-energy from the static calculation to $\text{Im}\Sigma (i\omega_n)=A(i\omega_n)^\alpha$ 
yields $\alpha \approx 0.75$, which is much closer to Fermi-liquid behavior. 
In this sense, the dynamic-$U$ calculation, by increasing the interaction effects, shifts the material closer to the 
spin freezing transition, and the proximity to this transition line results in a sensitive  
dependence of simulation results on parameters such as temperature and doping level. 

Figure~\ref{sigma}b shows the low-energy part of the self-energy of BaFe$_2$As$_2$ at $T=145$~K for different dopings. 
Increasing the hole doping leads to a large scattering rate 
while reducing the doping reduces the slope of $-\text{Im}\Sigma$ and hence the band renormalization. In fact, for undoped and electron
doped materials we find a Fermi-liquid behavior at the lowest temperatures, with mass renormalization factors 
$2.6$ (undoped) and $1.8$ (electron doping $x=0.2$ per Fe), respectively.

Our results on the NFL behavior are summarized in Fig.~\ref{sigma}c, which sketches the phase-diagram in 
the space of temperature and doping. The incoherent metal regime is shown in yellow, and the Fermi-liquid region in blue. The dotted line
is defined by the power-law exponent $\alpha=0.5$ (with simultaneous onset of static magnetic moments) and represents the boundary of the spin-frozen region in the sense of Ref.~\onlinecite{Werner08}. The dashed line marks the temperature and doping level below which Fermi-liquid behavior is recovered. 

\begin{figure}[h]
\centering
\includegraphics[angle=0, width=\linewidth]{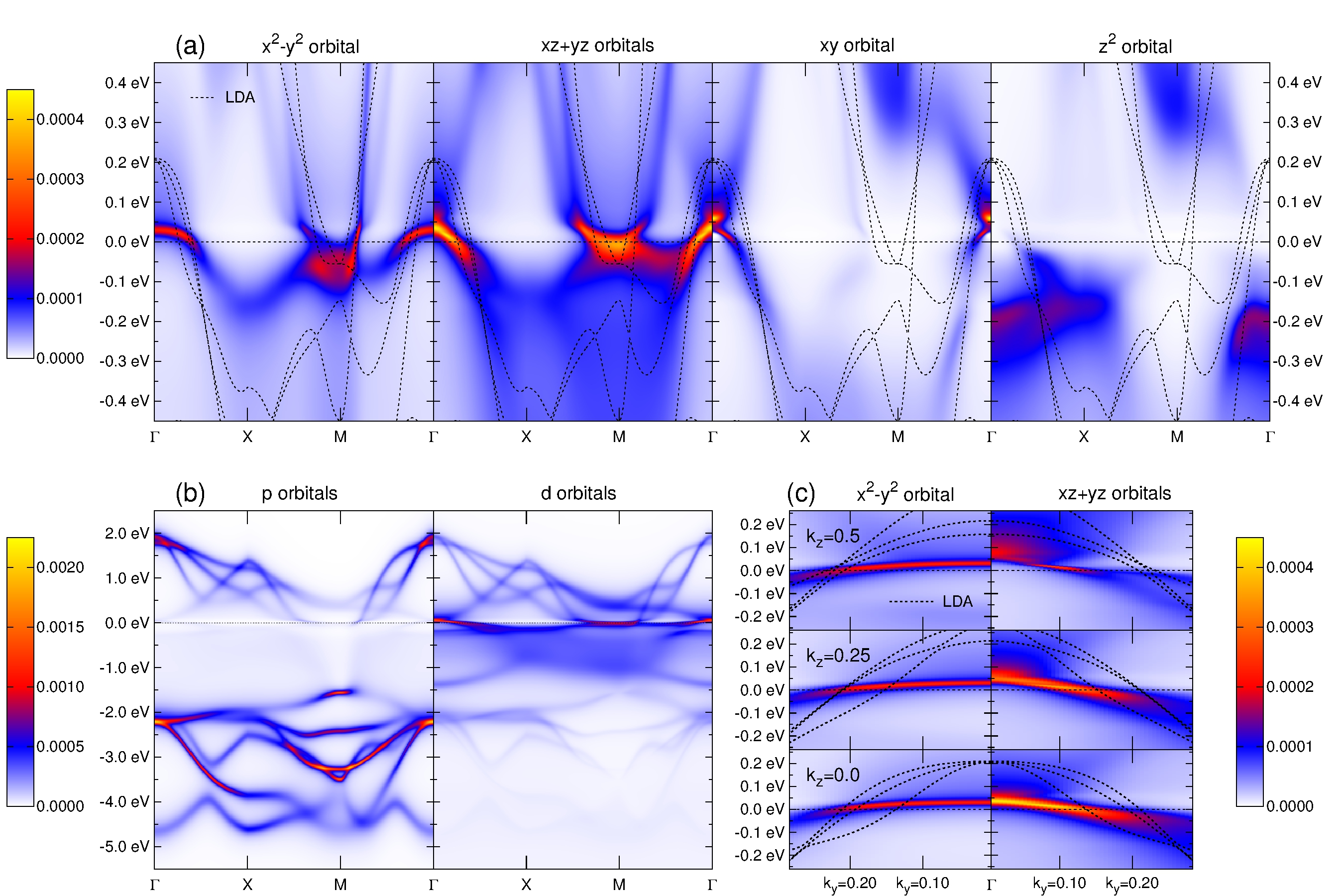}
\caption{Orbitally resolved spectral functions for optimally doped BaFe$_2$As$_2$ at $T=145$ K. Panel (a) shows the spectral functions along the path $\Gamma$-X-M-$\Gamma$ and compares the DMFT result to LDA calculations (dashed). The bands are strongly renormalized near the Fermi energy. Panel (b) shows the total $p$- and $d$-contribution to the spectral function in a wide energy window. The $p$-states hybridize with the $d$-states mainly in the unoccupied part of the spectrum. This is because it is precisely this hybridization that pushes the $d$-states up in energy. Panel (c) shows a cut through the hole pocket near $\Gamma$, for different values of $k_z$, and resolved into $x^2-y^2$ and $xy+yz$ contributions. Dotted lines indicate the LDA bands.
}
\label{pdcontribution}
\end{figure}

\noindent \textbf{Momentum-resolved spectral functions and comparison to experiment}

Spectral functions for optimally doped BaFe$_2$As$_2$ are shown in Fig.~\ref{pdcontribution}. 
The orbitally resolved $d$-electron spectral functions along the path $\Gamma$-X-M-$\Gamma$ are plotted in a narrow energy range in panel (a). A comparison with the original LDA band structure (dotted lines) shows the strong band renormalization at low energies, a consequence of the ``square root" behavior of the self-energy near the spin freezing transition. 
We also see that the $z^2$ orbital has no weight at the Fermi energy, which at first sight might suggest a simpler model for BaFe$_2$As$_2$, involving less $d$-orbitals. 
However, due to the sensitive doping dependence of the spin-freezing phenomenon, such models may have properties which are very different from those of our five-orbital model. The spin freezing transition produces dramatic effects in the metallic phase 
{\it away from half-filling}, where it leads to strong band renormalizations, even though the material is 
not close to any Mott insulating phase. In the five-orbital description, optimally doped BaFe$_2$As$_2$, with $\sim 5.8$ electrons per Fe,
falls into this NFL region (far away from the $6$-electron Mott phase). On the other hand, in a nearly half-filled four-orbital model, 
NFL effects comparable to those demonstrated in Fig.~\ref{sigma} are expected to appear only in the vicinity of the Mott insulating phase. 

Panel (b) plots the total $p$- and $d$-contribution to the spectral function along the same path. Remarkably, the $p$-bands (energy range [$-5$~eV:$-2$~eV]) overlap with the $d$-bands (energy range [$-2$~eV:$2$~eV]) mainly in the unoccupied part of the spectrum. 
This property, which can already be seen at the LDA level \cite{Andersen}, is amplified in the DMFT description,
since the empty $d$-states are much better defined than
the filled states whose spectra are smeared out by the self-energy effects described above.  
It may lead to an additional asymmetry between electron and hole doping, 
beyond the strong doping asymmetry implied by the spin-freezing phenomenon. 

A close-up view of the hole-pocket near $\Gamma$ is shown for several values of $k_z$ in Fig.~\ref{pdcontribution}c. Dashed lines again indicate the LDA bands. While the renormalization of these bands is very large, 
the $k_z$-dispersion of the Fermi surface is found to be weak in the DMFT description. 
The outer Fermi surface has predominantly $x^2-y^2$ character and the inner Fermi surface $xz+yz$ character.

\begin{figure}[h]
\centering
\includegraphics[angle=0, width=\linewidth]{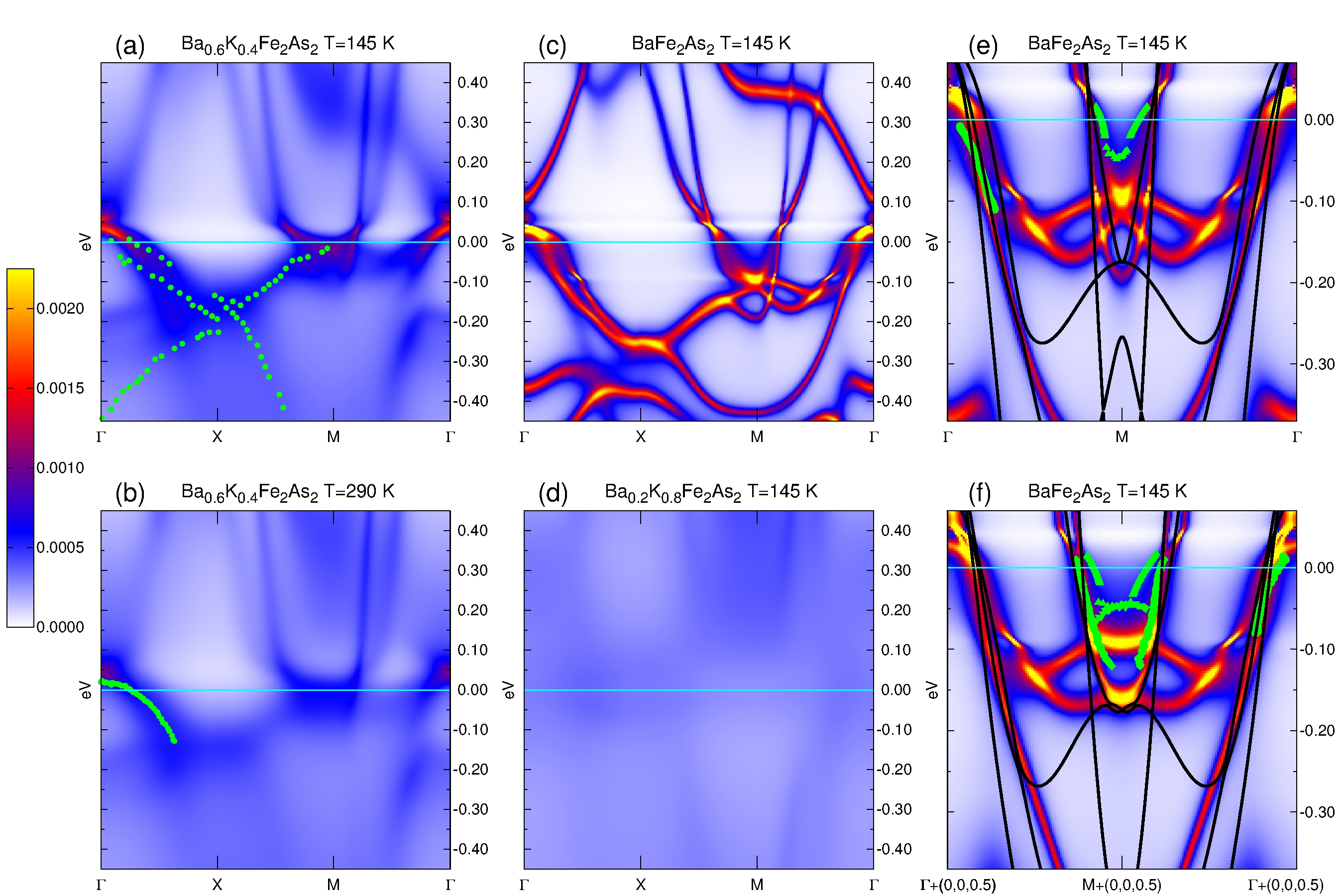}
\caption{Doping and temperature dependence of the low-energy spectral functions, and comparison to photoemission experiments. Panels (c) and (f) show the result for optimal doping ($x=0.2$) and compare the renormalized bandstructure to photoemission data taken from Ref.~\onlinecite{Ding08}. We plot the result for two temperatures, $T=145$ K (top) and $T=290$ K (bottom), to illustrate the smearing of the bands at elevated temperature. Panels (b) and (e) illustrate the doping dependence. The undoped compound is in the moderately correlated Fermi-liquid regime, while the overdoped compound ($x=0.4$) is in the incoherent metal regime, where the scattering off static moments washes out the bands. Panels (a) and (d) compare simulation results for the undoped compound at $T=145$ K to photoemission data from Ref.~\onlinecite{brouet-arxiv} and the LDA band structure (lines).}
\label{low_energy_spectra}
\end{figure}

A comparison of low-energy momentum-resolved spectra to photoemission data is presented 
in Fig.~\ref{low_energy_spectra}. 
Panels (a) and (b) show 
the spectral function along the path $\Gamma$-X-M-$\Gamma$, for optimal hole doping $x=0.2$, $T=145$ K (top) 
and $T=290$ K (bottom). Angle resolved photoemission data from Ref.~\onlinecite{Ding08} are indicated by green dots. 
(The photoemission spectra were measured at $T=50$ K and $T=145$ K, respectively).  
The agreement near the $\Gamma$, X and M points is remarkably 
good, which shows that the very strong low-energy renormalization
implied by the NFL self-energy is consistent with experiment. 
Increasing the temperature leads to a smearing of the bands, a result of the temperature dependent intercept 
of $\text{Im}\Sigma$ shown in the inset of Fig.~\ref{sigma}a.  
Panels (c) and (d) illustrate the strong doping dependence of the spectra. In panel (c), we show the result for the undoped compound at $T=145$ K, which is in the Fermi-liquid region of the schematic phase diagram (Fig.~\ref{sigma}c). Indeed, our calculation produces well-defined bands with a modest renormalization consistent with the factor $2.6$ extracted from the slope of the Matsubara axis self-energy. In panel (d), we plot the spectral function of the overdoped sample ($x=0.4$), which falls into the incoherent metal phase characterized by a non-zero intercept of $\text{Im}\Sigma$. As a result, the bands are smeared out. We also note 
that the electron pocket at M has disappeared. These theoretical predictions are consistent with measurements on the end member of the series, KFe$_2$As$_2$, reported in Refs.~\onlinecite{Sato2009,Yoshida2010}. Panels (e) and (f) of Fig.~\ref{low_energy_spectra} compare the low-energy band structure in the undoped compound at $T=145$ K to photoemission data taken from Ref.~\onlinecite{brouet-arxiv} (green dots) and the LDA bands (black lines). In this case, our calculation underestimates the band renormalization somewhat. However, given the large differences in the band renormalizations between the optimally doped and undoped compounds, we consider the agreement with the theoretical calculation satisfactory. We note that the discrepancy with the experiment is less than 0.05 eV, and thus much smaller than in any previous theoretical work. 

The spin-spin correlation function is known to exhibit an unusually slow (imaginary time) decay near the 
spin freezing-transition, while there is no particular anomaly in the orbital correlation functions~\cite{Werner08}. 
Our observation that the maximum $T_c$ in BaFe$_2$As$_2$ is reached in the vicinity of the spin-freezing transition
may therefore suggest that the spin degrees of freedom, rather than the orbital degrees
of freedom play an important role in the mechanism leading to high temperature
superconductivity.

\newpage

\begin{small}
\section{Methods}
Our scheme can be viewed as an extension
of the combined LDA+DMFT method 
\cite{Anisimov97,Lichtenstein98}
to dynamical interactions, or as promoting
extended dynamical mean field theory
(EDMFT) \cite{Si96, Kajueter_thesis} to a realistic scheme
by combining it with input from electronic
structure calculations within the LDA and
constrained RPA frameworks. Alternatively, our method can be considered
as an approximation to a full GW+DMFT calculation
\cite{Biermann03}. The present
scheme simplifies the general formulation
in so far as only a local self-energy
is calculated (as in DMFT) and 
two-particle quantities are calculated at
a non-selfconsistent level. Thereby 
practical calculations become feasible even
for complex multiband systems such as the iron
pnictide compounds.

In a Hamiltonian formulation, we can write the 
multi-orbital model with dynamically screened interactions as 
\begin{eqnarray}
H &=&\sum_{\{im\sigma \}}
(H_{im,i^{\prime }m^{\prime }}^\text{LDA}
- H_{im,i^{\prime }m^{\prime }}^\text{double~counting})
a_{im\sigma}^\dagger a_{i^{\prime }m^{\prime }\sigma }  \label{aalc1}  \nonumber \\
&+&\frac 12\sum_{imm^{\prime}\sigma \atop \text{\,(correl.~orb.)}}V_{mm^\prime}^i n_{im\sigma
}n_{im^{\prime }-\sigma }  \nonumber \\
&+&\frac 12\sum_{im\neq m^{\prime}\sigma \atop \text{\,(correl.~orb.)}
}(V_{mm^{\prime }}^i-J_{mm^{\prime
}}^i)n_{im\sigma }n_{im^{\prime }\sigma }  \nonumber
\\
&+& \sum_{i}\int d\omega \Big[\lambda_{i \omega}(b_{i\omega}^{\dagger} + b_{i\omega}) 
\sum_{m \sigma} n_{i m \sigma}+\omega b_{i\omega}^{\dagger}b_{i\omega}\Big].
\end{eqnarray}
It consists of a multiorbital Hubbard-Hamiltonian
of the form usually treated within LDA+DMFT \cite{Lichtenstein98},
albeit with the bare Coulomb interaction $V$ entering
the Hubbard-terms, and an additional bosonic
Hamiltonian with bosonic modes coupling to the
total electronic occupations of the atomic
sites. Here, $a_{im \sigma}^\dagger$ creates an electron on atom $i$
in orbital $m$ with spin $\sigma$ and 
$n_{i m \sigma}$ is the number operator for such 
electrons. 
For computational reasons, we have restriced ourselves to
density-density interactions.
The bosonic part of the Hamiltonian
describes the coupling of the electronic degrees 
of freedom (via the total charge $N_i=\sum_{m \sigma} n_{i m \sigma}$ on site $i$ and some coupling constant $\lambda_{i\omega}$)
to bosonic modes. These bosonic modes represent both, collective
plasmon oscillations and modes parametrizing one-particle
screening processes. 

We solve the multi-orbital lattice problem using dynamical 
mean field theory (DMFT) \cite{Georges96}, which maps it to a self-consistent 
solution of a five-orbital quantum impurity model. 
This local approximation, and the integration over the bosonic degrees of freedom, 
leads to an action of the form
\begin{eqnarray}
S_\text{imp}&=& \int \int d\tau d\tau^{\prime}
a_{m\sigma}^\dagger(\tau) \mathcal{G}_{0 m m^{\prime} \sigma}^{-1}(\tau - \tau^{\prime}) 
a_{m^{\prime }\sigma } (\tau^{\prime})
\nonumber
\\
&+&\frac 12\int d\tau \sum_{mm^{\prime}\sigma \atop \text{\,(correl.~orb.)}}
V_{mm^{\prime }} n_{m\sigma}(\tau)n_{m^{\prime }-\sigma }(\tau)  \nonumber \\
&+&\frac 12\int d\tau\sum_{m\neq m^{\prime}\sigma \atop \text{\,(correl.~orb.)}}
(V_{mm^{\prime }}-J_{mm^{\prime}})n_{m\sigma }(\tau)n_{m^{\prime }\sigma }(\tau)  
\nonumber
\\
&+&
\frac{1}{2}\int d\tau d\tau' N(\tau) U_\text{retarded}(\tau-\tau')N(\tau')
, 
\end{eqnarray}
with $-\text{Im} U_\text{retarded}(\omega)=\pi\lambda_\omega^2$, 
and the partially screened interaction $U_0=V-2\int d\omega \frac{\lambda_{\omega}^2}{\omega} $.

The frequency dependent (or retarded) $U$ 
can be viewed as a systematically downfolded
interaction, stemming from a static Hamiltonian including
all -- even high-energy -- degrees of freedom, in the
sense of Ref.~\onlinecite{aryasetiawan04}. 
In practice, we calculate it from the constrained RPA
method as described in the Supplementary Material.
The impurity model (with frequency dependent interactions) is solved using a Monte Carlo method based on a stochastic expansion of the partition function in the impurity-bath hybridization \cite{Werner06, Werner10}. Details can be found in the Supplementary Material.  

\bigskip

\section{Acknowledgements}
\acknowledgements We thank M.~Aichhorn, H.~Aoki, R.~Arita, V.~Brouet, H.~Ding, T.~Qian, and L.~Vaugier
for stimulating discussions. The CTQMC
calculations were run on the Brutus cluster at ETH Zurich using a code based
on ALPS \cite{ALPS}. We thank E. Gull for providing the code for the DMFT-selfconsistency loop.  
This work was supported by the Swiss National Science
Foundation (Grant PP002-118866), the G-COE program of MEXT (G-03),
the French ANR under project CORRELMAT, and IDRIS/GENCI Orsay
under project 20111393.  
We also acknowledge the hospitality of the KITP
Santa Barbara, where this work was initiated. \\


\end{small}


\newpage

\noindent \textbf{Supplementary Information}

These supplementary notes describe the methods used in our ab-initio simulation of BaFe$_2$As$_2$ with dynamically screened interactions. 

\subsection{Frequency dependent interaction from constrained RPA}

\begin{figure}[ht]
\centering
\includegraphics[width=0.5\linewidth]{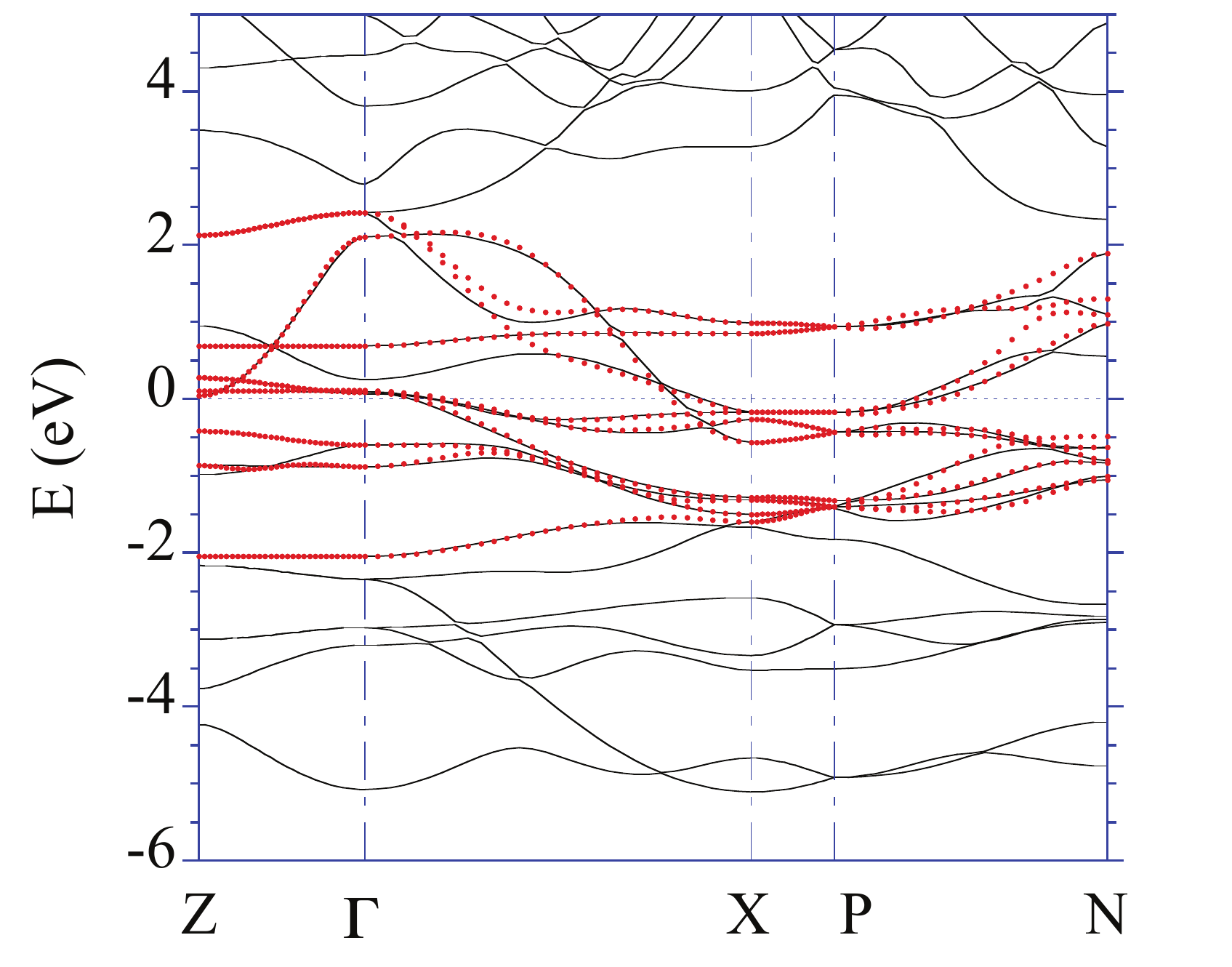}
\caption{
Electronic band structure of BaFe$_2$As$_2$.  
The black lines show the LDA band structure, whereas 
the red dots are interpolated bands in the $d$ subspace 
obtained using the maximally localized Wannier function procedure.
}
\label{band}
\end{figure}

In the constrained RPA (cRPA) method \cite{aryasetiawan04,aryasetiawan06} 
we first define a one-particle subspace $\left\{ \psi_{d}\right\} $ 
of the low-energy space, which we call the ``$d$ subspace", 
and label the rest of the Hilbert space by $\left\{\psi _{r}\right\} $ (``$r$ subspace"). 
In the present case, we choose 10 states having strong Fe-3$d$ character as the $d$ subspace (see Fig.~\ref{band}). 
We define $P_{d}({\bf r},{\bf r'};\omega)$ to be
the polarization within the $d$ subspace and $P({\bf r},{\bf r'};\omega)$ as the total polarization.
The rest of the polarization $P_{r}=P-P_{d}$ is \emph{not}\textbf{\ }the
same as the polarization of the $r$ subspace alone because it contains
polarization arising from transitions between the $d$ and $r$ subspaces. 
The physical idea behind the cRPA method is that the Hubbard \emph{U} defined as
an effective interaction of the electrons in the low-energy effective theory must be
such that when it is screened by the polarization of the low-energy states, it should be
equal to the fully screened interaction $W$ of the whole system. 
We thus define the {\it partially screened} Coulomb interaction $W_r({\bf r}, {\bf r'};\omega)$ by 
\begin{equation}
W_r(\omega )=[1 - V P_{r}(\omega )]^{-1} V,  \label{Wr} 
\end{equation}%
with $V$ denoting the bare Coulomb interaction.
The $d$ states are entangled with others in the present system. 
In order to distinguish between $P_d$ and $P_r$, we apply a disentangling procedure \cite{miyake09}. 

The Hamiltonian for our DMFT calculations contains 
ten bands around the Fermi level with dominantly Fe-$d$ character 
and also six bands (located at $-6$ to $-2$ eV) mainly coming from the As-$p$ bands. 
The Fe-$d$ orbitals are treated as correlated, whereas  the As-$p$ orbitals 
are assumed to be non-interacting.  
The Hubbard ${U}$ interactions for the Fe-$d$ orbitals are defined 
as matrix elements of $W_r$ in the maximally localized Wannier function 
(MLWF) \cite{marzari97,souza01} basis for the Fe-$d$ orbitals. 
More technical details on the estimation of $U$ can be found in Refs.~\onlinecite{miyake08} and \onlinecite{aichhorn09}. 
\begin{figure}[hb]
\centering
\includegraphics[angle=0, width=0.5\linewidth]{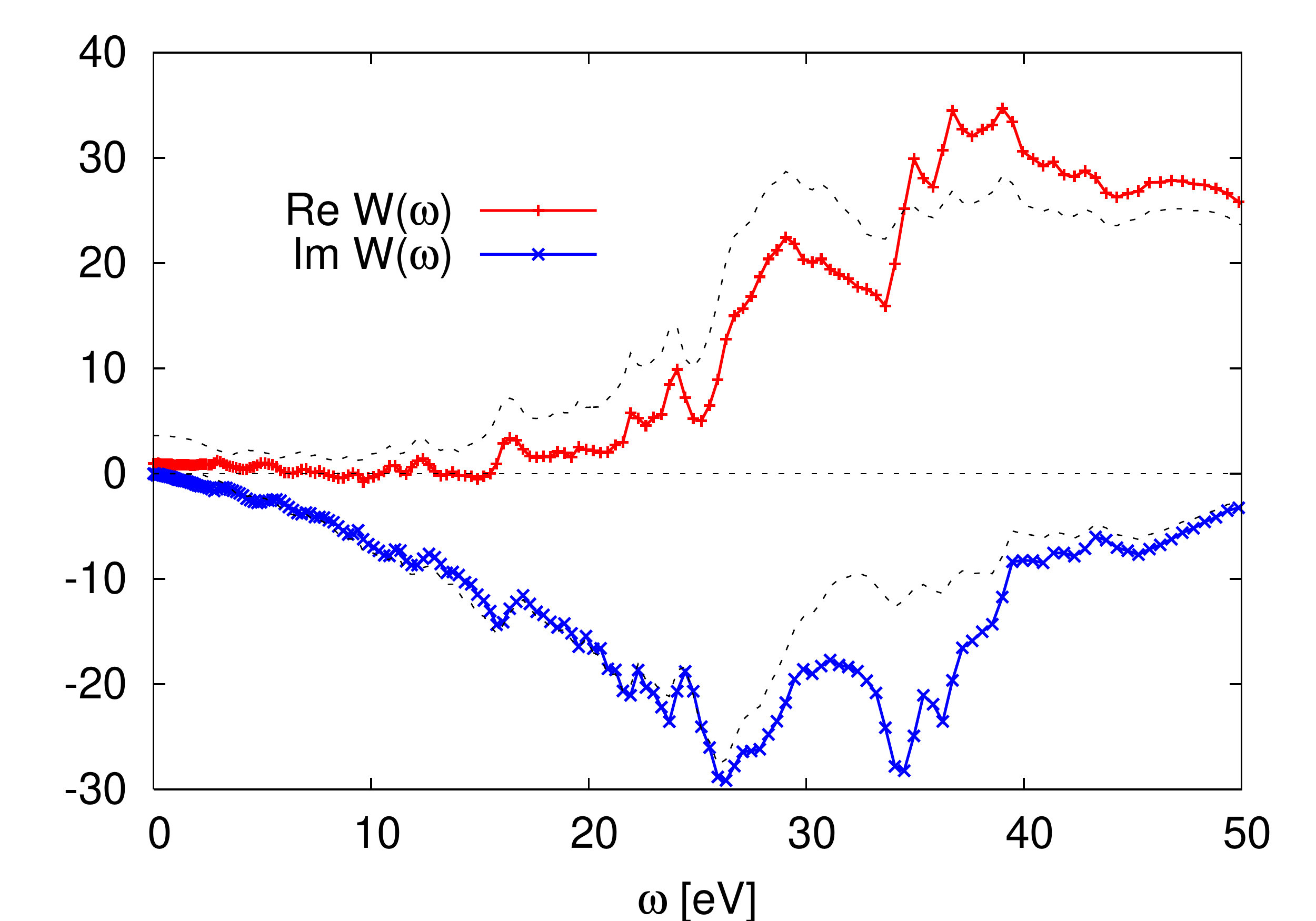}
\caption{
Frequency dependence of the fully screened Coulomb interaction $W(\omega)$ for BaFe$_2$As$_2$.  
The dashed lines show the result for the partially screened Coulomb interaction $U(\omega)$.
}
\label{wfreq}
\end{figure}

The frequency dependence of the partially screened interaction $U(\omega)$ is plotted in 
Fig.~1a of the main text. To illustrate the difference to the fully screened 
interaction $W(\omega )=[1-VP(\omega )]^{-1}V$, we plot $W(\omega)$ in Fig.~\ref{wfreq}. Note that 
$\text{Re}U(\omega=0)=3.61$ eV, while $\text{Re} W(\omega=0)=0.975$ eV. 
The Hund coupling parameter $J$ is approximately $0.8$ eV and shows little frequency dependence. 

The frequency dependence of the Hubbard \emph{U} reflects retardation
effects arising from the $r$ subspace that has been eliminated in the low-energy 
effective action. 
The derived action contains $p$ and $d$ orbitals with 
the onsite $U(\omega)$ term on the $d$ orbitals. 
In the $d$-manifold with orbitals ($xy$, $yz$, $z^2$, $zx$, $x^2-y^2$), the static values of the interaction for same and opposite spin are 
\begin{eqnarray}
U^{\sigma\bar\sigma}_{mn}(J=0.8)&=&
\left(
\begin{tabular}{lllll}
 3.61  & 2.38  & 2.20  & 2.38  & 2.93 \\
 2.38  & 3.61  & 2.74  & 2.38  & 2.38 \\
 2.20  & 2.74  & 3.61  & 2.74  & 2.20 \\
 2.38  & 2.38  & 2.74  & 3.61  & 2.38 \\
 2.93  & 2.38  & 2.20  & 2.38  & 3.61
\end{tabular}
\right),\label{opposite_spin}\\
U^{\sigma \sigma}_{mn}(J=0.8)&=&
\left(
\begin{tabular}{lllll}
 0.00  & 1.76  & 1.49 & 1.76 & 2.58 \\
 1.76  & 0.00  & 2.31 & 1.76 & 1.76 \\
 1.49  & 2.31  & 0.00 & 2.31 & 1.49 \\
 1.76  & 1.76 & 2.31 & 0.00 & 1.76 \\
 2.58 & 1.76 & 1.49 & 1.76 & 0.00  
\end{tabular}
\right), \label{same_spin}
\end{eqnarray}
and we take the frequency dependence shown in Fig.~1a of the main text for all components. In Eqs.~(\ref{opposite_spin}) and (\ref{same_spin}) we have symmetrized the matrix elements obtained from cRPA, by computing the Slater paramters $F^0$, $F^2$ and $F^4$ from the averaged $\bar U^{\sigma\bar\sigma}=2.70$~eV and $J=0.8$~eV, and reconstructing the intra- and inter-orbital interactions using the procedure described in Ref.~\onlinecite{Lichtenstein95}. This symmetrization procedure allows us to use an orbital independent double-counting term. 

Approximating the Coulomb interaction by the density-density terms leads to an underestimation of the spin fluctuations or, equivalently, to an overestimation of the role of the Hund coupling $J$. We found, in agreement with previous studies on other iron pnictides \cite{Haule_njp, Ishida10} that the simulation results depend very sensitively on the choice of the parameter $J$. To obtain meaningful results for the low-energy spectral function, we thus reduced the value of the Hund coupling to $J=0.675$, which (using $\bar U^{\sigma\bar\sigma}=2.70$~eV) gives the following interaction matrices:
\begin{eqnarray}
U^{\sigma\bar\sigma}_{mn}(J=0.675)&=&
\left(
\begin{tabular}{lllll}
 3.61 & 2.57 & 2.42 & 2.57 & 3.03 \\
 2.57 & 3.61 & 2.88 & 2.57 & 2.57 \\
 2.42 & 2.88 & 3.61 & 2.88 & 2.42 \\
 2.57 & 2.57 & 2.88 & 3.61 & 2.57 \\
 3.03 & 2.57 & 2.42 & 2.57 & 3.61 
\end{tabular}
\right),\label{opposite_spin0.675}\\
U^{\sigma \sigma}_{mn}(J=0.675)&=&
\left(
\begin{tabular}{lllll}
 0.00 & 2.05 & 1.82 & 2.05 & 2.74 \\
 2.05 & 0.00 & 2.51 & 2.05 & 2.05 \\
 1.82 & 2.51 & 0.00 & 2.51 & 1.82 \\
 2.05 & 2.05 & 2.51 & 0.00 & 2.05 \\
 2.74 & 2.05 & 1.82 & 2.05 & 0.00
\end{tabular}
\right). \label{same_spin0.675}
\end{eqnarray}

\subsection{Dynamical mean field (DMFT) calculation with frequency dependent interaction}

The Coulomb matrix elements (\ref{same_spin0.675}) and (\ref{opposite_spin0.675}), their frequency dependence (encoded by $\text{Im} U(\omega)$) and the Hamiltonian matrix $H_k$ for the $p$ and $d$ bands in the Wannier basis are the input of the DMFT calculation. DMFT neglects the momentum dependence of the self-energy and replaces the lattice problem by a 5-orbital impurity model with frequency dependent local interactions, and a self-consistency procedure involving $H_k$, which fixes the hybridization functions \cite{Kotliar06}. 
(How a momentum-dependent self-energy affects these multi-orbital calculations is an interesting open problem.)

Since the LDA bandstructure ($H_k$) already captures some correlation effects in the $d$-orbitals, we modify the self-energy $\Sigma_d$ by an orbital independent shift (double-counting correction) of 
\begin{equation}
E_{DC}=F^0 (n_d-1/2)-J(n_d/2 - 1/2),
\end{equation}
with $n_d$ the self-consistently computed number of $d$ electrons. 
This double counting was found in Ref.~\onlinecite{Aichhorn_arxiv} to yield the best agreement between charge-selfconsistent and non-selfconsistent calculations. Note that the static value of the interaction appears in $E_{DC}$. This is because the addition of bosonic modes in a dynamically screened model requires the introduction of additional double counting terms for these modes, which eliminate the bare interaction from the double counting formula. 

The new feature, compared to previous LDA+DMFT simulations, is the treatment of the full frequency dependence of the interaction. We employ the method developed in Refs.~\onlinecite{Werner07, Werner10}, which is based on the hybridization expansion approach \cite{Werner06}. This technique can be easily generalized to multi-orbital systems. We define the bosonic factor $b(\omega',\tau)=\cosh\big[\big(\tau-\frac{\beta}{2}\big)\omega'\big]/\sinh\big[\frac{\omega'\beta}{2}\big]$ and the function (valid for $0\le \tau \le \beta$)
\begin{eqnarray}
K(\tau)&=&\int_0^\infty \frac{d\omega'}{\pi}\frac{\text{Im}U(\omega')}{\omega'^2}[b(\omega',\tau)-b(\omega',0)],
\label{k_definition}
\end{eqnarray}
which (up to the sign) corresponds to the twice integrated nonlocal interaction. 
The frequency dependence of $U$ then enters the hybridization expansion calculation in the form of a non-local interaction between each pair of creation and/or annihilation operators (irrespective of orbital) of the form $w_{ij}=\exp[s_i s_j K(\tau_i-\tau_j)]$, where $\tau_i>\tau_j$ are the positions of the two hybridization events on the imaginary time interval and $s=1$ for creation operators and $-1$ for annihilation operators. 
Apart from these modifications, the impurity calculation proceeds as usual (with the static value of $U$ for the evaluation of the interaction contribution to the Monte Carlo weight), via random insertions and removals of pairs of hybridization operators. For models with density-density interaction, this method is highly efficient. At the lowest temperature, $T=145$ K, we used about 10 CPU hours per iteration.  

\subsection{Analytical continuation}
\label{analyticalcontinuation}

The DMFT calculation yields an imaginary-time Green function which contains all the dynamic features encoded in the retarded interaction $U(\omega)$. However, to analyze the effects of the retarted interaction on the spectral function, an inversion procedure is required which allows us to go from the imaginary time to the real time domain. The stochastic noise in the CTQMC data makes this problem extremely difficult for standard Maximum entropy methods if one aims at resolving intermediate-to-high energy features like the satellites discussed in the main text. In Ref.~\onlinecite{dyn_anal_cont}, a new analytical continuation procedure has been proposed which is based on the exact atomic limit properties of quantum impurity problems with retarded interactions. In that limit, the exact Green function becomes the product of a purely static (local in time) interacting part, $G_\textrm{static}(\tau)$, and the factor $B(\tau)=\exp[-K(\tau)]$ with bosonic symmetry containing the full retarded tail of the interaction $U$. $G_\textrm{static}$ and $B$ are analytically known, and $B$ is responsible for both the low-energy renormalization of the Green function, and the satellites resulting from the screening processeses embedded in $U$. The information encoded in the $B$ factor gives the correct asymptotics and intermediate-to-high energy properties even away from the atomic limit, as the hybridization affects mainly the low-energy part of the spectral function. 

Motivated by these considerations we introduce an auxiliary Green function $G_\textrm{auxiliary}$ satisfying
\begin{equation}
G(\tau) = G_\textrm{auxiliary}(\tau) B(\tau).
\label{aux_definition}
\end{equation}
$G_\textrm{auxiliary}$ describes mainly the low-energy features of the full Green function, thanks to the energy scale separation in the spectrum. Therefore, the standard maximum entropy method can be applied reliably to compute the spectral function $\rho_\textrm{aux}$ of $G_\textrm{auxiliary}$, while $\rho_B$ - the spectral representation of the Bose factor $B$ - can be obtained via the numerical integration of Eq.~(\ref{k_definition}) at any desired accuracy. The full spectral function is obtained from the integral 

\begin{equation}
\rho(\omega) = \int_{-\infty}^\infty \!\!\! \dd\epsilon ~ \rho_B(\epsilon) 
\frac{1+e^{-\beta\omega}}{(1 + e^{-\beta(\epsilon-\omega)})(1 - e^{-\beta\epsilon})} 
\rho_\textrm{aux}(\omega-\epsilon).
\label{conv}
\end{equation}

The bosonic spectral function $\rho_B$ for BaFe$_2$As$_2$ is shown in Fig. 1b of the main text. It 
essentially inherits the structures from $\text{Im} U(\omega)/\omega^2$. At zero temperature, Eq.~(\ref{conv}) reduces to the convolution of $\rho_B$ and $\rho_\text{aux}$. If $\rho_\text{aux}(\omega)$ has a sharp peak at $\omega\approx 0$, this convolution will produce a satellite at each of the energies corresponding to sharp features in $\rho_B$. 

To analytically continue the self-energy, $\Sigma(i\omega_n)$, we define an effective Green function $G^\Sigma(i\omega_n)=[i\omega_n-\mu_\text{eff}-\Sigma(i\omega_n)]^{-1}$ and apply the above procedure to $G^\Sigma(i\omega_n)$. From its spectral function $\rho^\Sigma(\omega)$, the calculation of $\Sigma(\omega)$ follows straightforwardly through the Kramers-Kronig relations.

\end{document}